# Site-Projected Thermal Conductivity: Application to Defects, Interfaces, and Homogeneously Disordered Materials

*Aashish Gautam, Yoon Gyu Lee, Chinonso Ugwumadu, Kishor Nepal, Serge Nakhmanson, and David A. Drabold\**

With the rapid advance of high-performance computing and electronic technologies, understanding thermal conductivity in materials has become increasingly important. This study presents a novel method: the site-projected thermal conductivity that quantitatively estimates the local (atomic) contribution to heat transport, leveraging the Green–Kubo thermal transport equations. The effectiveness of this approach on disordered and amorphous graphene, amorphous silicon, and grain boundaries in silicon–germanium alloys is demonstrated. Amorphous graphene reveals a percolation behavior for thermal transport. The results highlight the potential of the method to provide new insights into the thermal behavior of materials, offering a promising avenue for materials design and performance optimization.

## 1. Introduction

The investigation of thermal transport properties in materials is a rapidly evolving field of research.[1–4] A key challenge lies in identifying and characterizing thermally active sites, which is crucial for understanding and predicting thermal transport behavior. While spatially resolved models for electronic transport have been successfully developed and applied to a broad range of materials,[5–11] achieving a similarly comprehensive representation of thermally active sites remains an area of intense and ongoing research.

In this study, we present the site-projected thermal conductivity (SPTC) method, which offers an intuitive framework for identifying thermally active sites within materials in real space. We adopt the Allen and Feldman approach[12] to derive an expression for thermal conductivity using the Green–Kubo formalism for thermal transport.[13] The method is formulated for disordered systems including topological and chemical disorders, extended, and point defects. The SPTC method is applied to a range of crystalline, defective, and amorphous materials, including carbon and silicon allotropes. Additionally, we investigate the thermal properties of interfaces and grain boundaries (GBs) in paracrystalline silicon and silicon–germanium alloy systems. It is important to note that this method accounts solely for phonon contributions to heat transport, excluding electronic effects.

A. Gautam, Y. G. Lee, K. Nepal, D. A. Drabold
Department of Physics and Astronomy
Ohio University
Athens, OH 45701, USA
E-mail: drabold@ohio.edu

C. Ugwumadu
Physics of Condensed Matter and Complex Systems (T-4) Group
Los Alamos National Laboratory
Los Alamos, NM 87545, USA

S. Nakhmanson
Department of Physics
Department of Materials Science & Engineering
Institute of Materials Science
University of Connecticut
Storrs, CT 06269, USA

The ORCID identification number(s) for the author(s) of this article can be found under https://doi.org/10.1002/pssr.202400306.

## 2. Theory

The theory of Allen and Feldman,[12] tersely recounted here, is a method for computing and understanding thermal transport in materials in harmonic approximation (HA). Their starting point is Kubo's expression for the thermal conductivity tensor, $\kappa_{\mu\nu}$.

$$\kappa_{\mu\nu}(\omega) = \frac{V}{T} \int_0^{1/k_B T} d\tau \int_0^\infty dt e^{i(\omega+i\eta)t} \langle S_\mu(-i\hbar\tau) S_\nu(t) \rangle \qquad (1)$$

Here, $V$ and $T$ are the volume and the temperature of the system, $\mu$ and $\nu$ are the Cartesian components of the heat current operator $\mathbf{S}$, and $\langle \ldots \rangle$ is the current–current autocorrelation function. Hardy's form for $\mathbf{S}$ is (with atoms indexed by Latin indices and directions by Greek indices $\alpha, \beta$)

$$\mathbf{S} = -\frac{1}{2V} \sum_{i,j} \sum_{\alpha,\beta} (\mathbf{R}_i - \mathbf{R}_j) \frac{p_{i\alpha}}{m_i} \frac{\partial^2 U}{\partial q_i^\alpha \partial q_j^\beta} q_j^\beta \qquad (2)$$

where $\mathbf{R}_i$ in the position of $i^{th}$ atom in the unit cell and $q_i^\alpha$ is the $\alpha$-component of the displacement vector for $i^{th}$ atom. Invoking the usual HA, the potential energy, $U$, is given as

$$U = \frac{1}{2}\sum_{\gamma,\gamma'}\sum_{i,j}\sum_{\alpha,\beta}\phi_{i,j}^{\alpha\beta}(\gamma,\gamma')q_{\gamma i}^{\alpha}q_{\gamma' j}^{\beta} \quad (3)$$

$$\phi_{i,j}^{\alpha\beta}(\gamma,\gamma') = \frac{\partial^2 U}{\partial q_{\gamma i}^{\alpha} \partial q_{\gamma' j}^{\beta}} \quad (4)$$

Here, $\phi_{i,j}^{\alpha\beta}(\gamma,\gamma')$ is the force constant matrix, and $\gamma$ and $\gamma'$ represent the cells in the periodic system. From the classical equations of motion

$$\omega_m^{\mathbf{k}\,2} e_{i,\mathbf{k}}^{\alpha,m} = \sum_{\beta j} D_{ij}^{\alpha\beta}(\mathbf{k}) e_{j,\mathbf{k}}^{\beta,m} \quad (5)$$

$$D_{ij}^{\alpha\beta}(\mathbf{k}) = \sum_{\gamma}\frac{1}{\sqrt{m_i m_j}}\phi_{i,j}^{\alpha\beta}(0,\gamma)e^{i\mathbf{k}\cdot(\mathbf{R}_{j,\gamma}-\mathbf{R}_{i,0})} \quad (6)$$

Here, $e_{i,\mathbf{k}}^{\alpha,m}$ is the polarization of the vibrational mode $m$ at atom $i$ along $\alpha$ direction. $D_{ij}^{\alpha\beta}(\mathbf{k})$ is the dynamical matrix (DM), the Fourier transform of the force constant matrix. $\mathbf{R}_{j,\gamma}$ is the position of $j^{th}$ atom in the $\gamma^{th}$ cell. $\mathbf{k}$ is a wave vector that may be chosen at the center or corner of the Brillouin zone. For $N$ cells in a system with $l$ atoms per cell, labeled by $i = 1, 2 \ldots, l$, the $\mathbf{S}$ operator from Equation (2) is written as

$$\mathbf{S} = \sum_{m,n} \mathbf{S}_{mn} \hat{a}_m^{\dagger} \hat{a}_n \quad (7)$$

$$\mathbf{S}_{mn} = \frac{\hbar}{2V} \mathbf{v}_{mn}^{\mathbf{k}}(\omega_m^{\mathbf{k}} + \omega_n^{\mathbf{k}}) \quad (8)$$

Here, $m, n$ are the indices of the phonon modes; $\hat{a}_m^{\dagger}$ and $\hat{a}_n$ are the creation and annihilation operators of a phonon mode respectively; $\mathbf{v}_{mn}^{\mathbf{k}}$ is the group velocity vector of mode $m$ when $m = n$; $\omega_m^{\mathbf{k}}$ is the vibrational frequency of the mode $m$. Upon applying the HA results on Equation (1), a general expression for the real part of TC is obtained.[12]

$$\kappa_{\alpha\beta}(\omega) = \frac{\pi V}{T\hbar}\sum_{m,n}\left[\frac{\langle f_m\rangle - \langle f_n\rangle}{(\omega_m - \omega_n)}\right](S_{mn}^{\alpha})(S_{mn}^{\beta})^{\dagger}\delta(\omega_m^{\mathbf{k}} - \omega_n^{\mathbf{k}} - \omega) \quad (9)$$

where $f_m$ is the equilibrium occupation of the $m^{th}$ mode (i.e., the Bose–Einstein distribution) and $(S^{\alpha})_{mn}$ is the $\alpha^{th}$ component of $S_{mn}$. When $\omega \to 0$, due to the presence of the $\delta$ function, the term $\frac{\langle f_m\rangle - \langle f_n\rangle}{(\omega_m - \omega_n)}$ in Equation (9) simplifies to $-\partial\langle f_m\rangle/\partial\omega_m$. With these considerations, and using Equation (7) and (8), Allen and Feldman's expression for the TC is

$$\kappa = \sum_{m,n\neq m} \frac{-\pi\hbar}{48VT}\left[\frac{\partial\langle f_m\rangle}{\partial\omega_m^{\mathbf{k}}}\right](\omega_m^{\mathbf{k}} + \omega_n^{\mathbf{k}})^2 \frac{\delta(\omega_m^{\mathbf{k}} - \omega_n^{\mathbf{k}})}{\omega_m^{\mathbf{k}}\omega_n^{\mathbf{k}}}$$

$$\sum_{\eta}\sum_{\alpha,\beta}\sum_{\gamma,x,x'} e_{x,\mathbf{k}}^{\alpha,m} e_{x',\mathbf{k}}^{\beta,n}\frac{1}{\sqrt{m_x m_{x'}}}\phi_{x,x'}^{\alpha\beta}(0,\gamma)\left(R_{\gamma}^{\eta} + R_{xx'}^{\eta}\right) \quad (10)$$

$$\sum_{\alpha',\beta'}\sum_{\gamma',a,b} e_{a,\mathbf{k}}^{\alpha',m\dagger} e_{b,\mathbf{k}}^{\beta',n\dagger}\frac{1}{\sqrt{m_a m_b}}\phi_{a,b}^{\alpha'\beta'}(0,\gamma')\left(R_{\gamma'}^{\eta} + R_{ab}^{\eta}\right)$$

where $R_{\gamma}^{\eta}$ is the $\eta$ component of position vector for cell $\gamma$; $R_{xx'}^{\eta}$ is the $\eta$ component of displacement between atoms $x$ and $x'$.

The thermal conductivity in Equation (10) is taken as an average of the Cartesian components of the conductivity tensor. It is worth noting that in the Allen–Feldman method, the lattice is properly quantized and so the method is suitable for arbitrarily low temperatures.

### 2.1. The Site-Projected Thermal Conductivity

To this point, we have only recounted the work of Allen and Feldman. In this section, we seek to extract spatial information about thermal conduction within the Allen and Feldman picture. As we have shown for the electrical conductivity,[5] simple rearrangements of expressions for the conductivity provide spatial information about conduction activity. That is, any Kubo formula may be rearranged such that the target conductivity is a double sum over spatial points (atomic sites in this case) $x$. We show that an informative expression for a spatially resolved TC emerges as a result, providing the most obvious form for it in this article. Other forms are possible, and we are exploring general considerations for an optimal form for SPTC.

Rearranging summation order in Equation (10) and taking $\mathbf{k} = 0$, we can obtain an expression for $\kappa$ as a (double) sum over terms depending on atomic positions $x$ and $x'$ in the unit cell.

$$\kappa = \sum_{x,x'}\Xi(x,x') \quad (11)$$

where

$$\Xi(x,x') = \frac{\pi\hbar^2}{48k_B T^2 V}\frac{1}{\sqrt{m_{x'} m_x}}\sum_{\eta,\gamma}\left(R_{\gamma}^{\eta} + R_{xx'}^{\eta}\right)\sum_{m,n\neq m}\delta(\omega_m^0 - \omega_n^0)$$

$$\frac{(\omega_m^0 + \omega_n^0)^2}{\omega_m^0 \omega_n^0}\left(\frac{e^{\frac{\hbar\omega_m}{k_B T}}}{\left(e^{\frac{\hbar\omega_m}{k_B T}} - 1\right)^2}\right)\sum_{\alpha\beta}\phi_{x,x'}^{\alpha\beta}(0,\gamma)e_{x,0}^{\alpha,m}e_{x',0}^{\beta,n}$$

$$\sum_{\gamma'ab}\sum_{\alpha'\beta'}\frac{1}{\sqrt{m_a m_b}}\phi_{a,b}^{\alpha'\beta'}(0,\gamma')e_{a,0}^{\alpha',m\dagger}e_{b,0}^{\beta',n\dagger}\left(R_{\gamma'}^{\eta} + R_{ab}^{\eta}\right) \quad (12)$$

For $\omega = 0$, $\Xi$ is a Hermitian matrix, that is, $\Xi(x,x') = \Xi(x',x)$. In computing $\Xi$, care should be taken to exclude very small $\omega_i^0$. Equation (11) suggests a way to decompose the total TC into contributions depending on atomic position $x$ by conducting a summation of $\Xi(x,x')$ overall positions $x'$ in the supercell.

$$\zeta(x) = \sum_{x'}\Xi(x,x') \quad (13)$$

Here, $\zeta(x)$ can be named the SPTC, as it represents the contribution of an atom at site $x$ to the total TC of the system. Summing $\zeta(x)$ over atom index $x$ recovers the total TC $\kappa$.

$$\kappa = \sum_{x}\zeta(x) \quad (14)$$

## 3. Illustration of the Method

As a proof of concept, we conducted SPTC analysis on a range of different material systems, carbon, silicon, and silicon–germanium structures. These systems included crystalline graphene, with both single- and double-vacancy defects, as well as amorphous graphene, characterized by ring disorder. Additionally, we applied SPTC analysis to diamond–crystalline silicon, paracrystalline silicon, and silicon–germanium alloy systems, focusing on GBs and mass defects.

Molecular dynamics simulations were conducted using the large-scale atomic/molecular massively parallel simulator (LAMMPS).[14] Within LAMMPS, the machine learning Gaussian approximation potentials for carbon[15] and silicon[16] were employed for the carbon and pure silicon simulations. The Tersoff potential was used for the silicon–germanium structure.[17]

Periodic boundary conditions were applied for all the simulations. The conjugate gradient algorithm was used for energy minimization (and structural optimization) with an energy convergence threshold of $10^{-10}$ eV. For the DM calculations, the atoms were displaced by 0.05 Å in $x$–, $y$–, and $z$–directions, and the SPTC calculations were performed using Equation (13). The subsequent sections elaborate on the specific simulation protocols for the materials studied.

### 3.1. Carbon Structures

For the SPTC analysis of crystalline carbon systems, a structural model of crystalline graphene with surface dimensions of 29.4 × 34.03 Å was utilized, as illustrated in **Figure 1**a. To investigate the influence of vacancy defects on TC, we created models for 1) Stone–Wales defect[18] with pentagon–octagon–pentagon rings (henceforth referred to as SW-585) and 2) a single-vacancy defect. The SW-585 model was constructed by removing two atoms from the graphene sheet and performing a structural relaxation. Both vacancy models are presented in Figure 1b,c, respectively.

The initial configuration of the 2000-atom amorphous graphene model (aGr$_{2000}$) was generated by randomly distributing atoms within a cuboid measuring 72.30 × 72.30 × 4 Å. The simulation followed the methodology for amorphous graphite outlined in refs. [19,20], with a modification during the heating process. The system was simulated in the canonical ensemble (NVT), controlled by a Nosé–Hoover thermostat[21–23] at 2700 K. Movement was confined to the XY plane. Afterward, the $z$ dimension was extended by 3 Å (on both sides), and the structure was annealed at 300 K for 50 ps before being relaxed, resulting in the 3D undulating structure shown in Figure 1d. We note that the amorphous graphene constructed in this study is highly defective but serves its purpose in revealing the effects of defects on thermal conduction pathways in disordered graphene. Less-defective amorphous graphene can be obtained using the methods described in refs. [24,25].

### 3.2. Silicon Single-Vacancy Defect

A diamond-structure Si unit cell was sourced from the Materials Project,[26] and a 1000-atom, 5 × 5 × 5 supercell model was constructed from it and used as the base for creating different Si structures. The original cubic supercell, with a length of 27.21 Å and density ($\rho$) of 2.33 g cm$^{-3}$, was relaxed, exhibiting no significant change in its conformation. A single-vacancy defect, shown in **Figure 2**a, was created by removing an atom from the crystalline Si lattice, followed by another energy/structural relaxation. For the analysis of the temperature and frequency dependence of TC in amorphous silicon (aSi), a tetrahedral Si model with 512 atoms ($\rho = 2.34$ g cm$^{-3}$), obtained from ref. [27], was utilized.

### 3.3. Twin Grain Boundary in Silicon

A silicon GB model, denoted as $\Sigma3(112)/[1\bar{1}0]$, was obtained from ref. [28]. The cell dimensions are 9.46 × 15.45 × 100.60 Å$^3$, with a separation of ≈50.30 Å between the two GB planes, as shown in Figure 2b. The twin GB exhibits a misorientation characterized by a 60° rotation around the $[1\bar{1}0]$ axis, with one of the grains aligned along the (112) plane.

### 3.4. Alloy-Chemical Disorder (Si$_{1-x}$Ge$_x$ Structure)

To construct Si$_{1-x}$Ge$_x$ structures, for $x = 0.25$, 0.5, and 0.75, the silicon atoms in the diamond Si supercell discussed in Section 3.2 were randomly replaced with germanium atoms followed by conjugate gradient relaxation. Figure 2c shows a representation of the resulting structure. The Si$_{0.75}$Ge$_{0.25}$ composition was selected to investigate the effects of mass disorder on the thermal transport behavior.

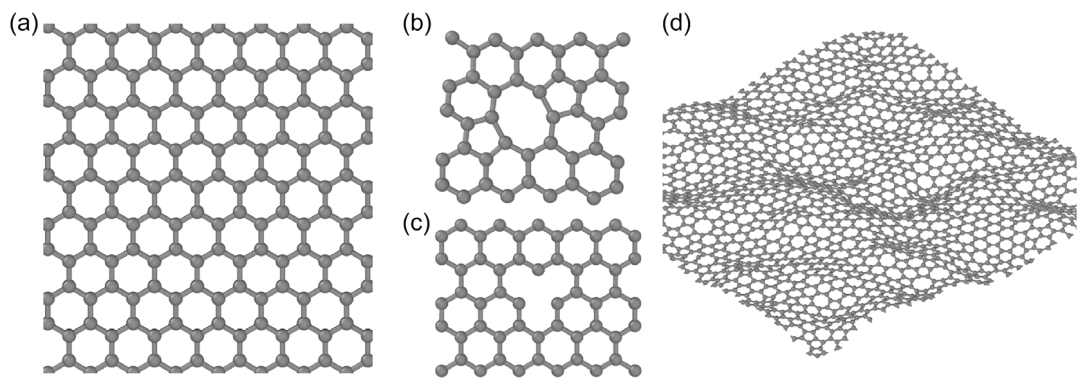

**Figure 1.** Representations of the carbon-based models: a) crystalline graphene, b) 585 Stone–Wales defect, c) single-vacancy defect, and d) amorphous graphene.

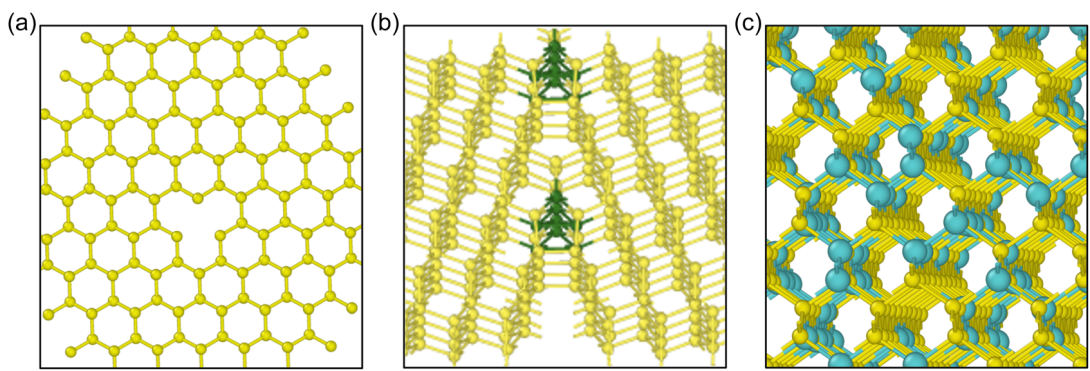

**Figure 2.** Representations of the silicon models (yellow spheres): a) single-vacancy defect in diamond Si, b) Σ3(112)/[1$\bar{1}$0] twin GB, with fivefold coordinated Si atoms at the boundary shown as dark green spheres, and c) silicon–germanium structure, where germanium atoms (cyan spheres) are randomly substituted for Si atoms in the diamond Si lattice.

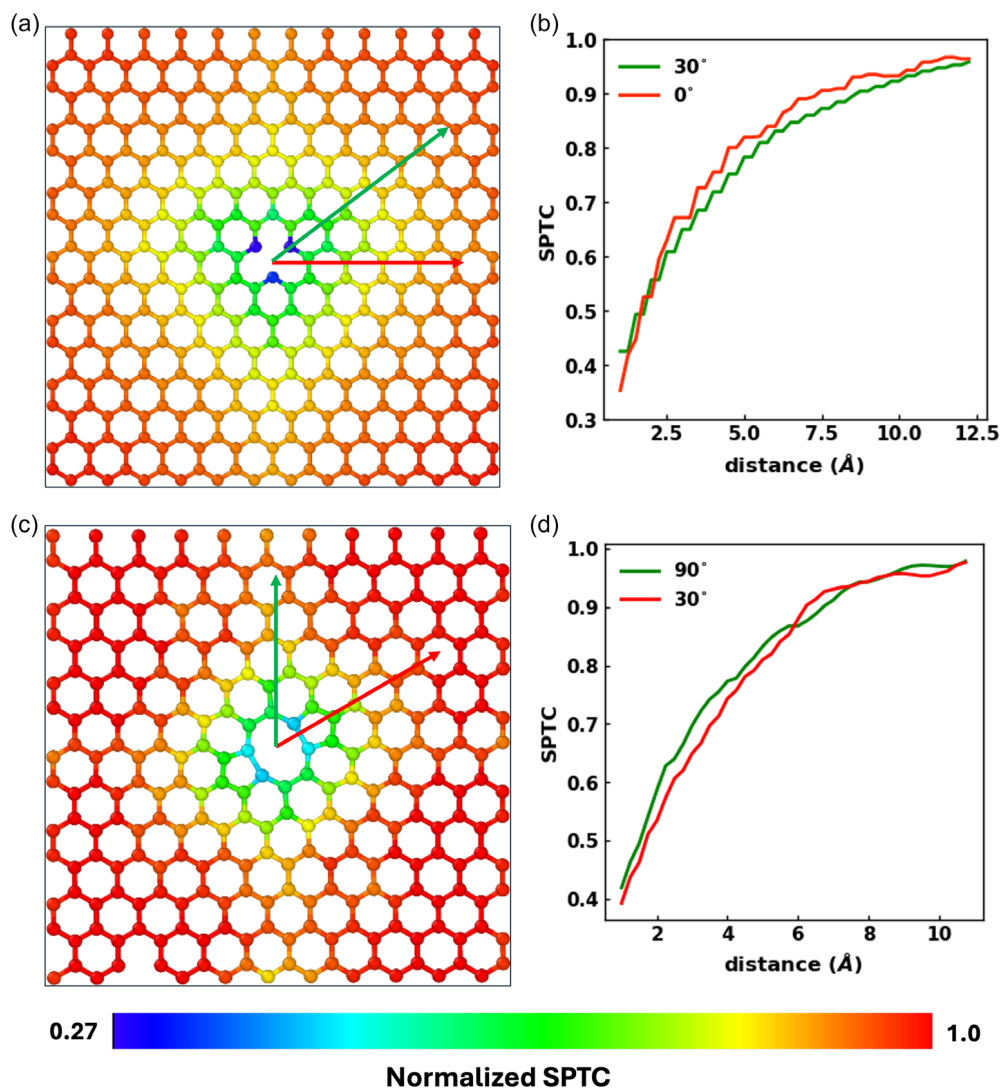

**Figure 3.** a) Plot of the normalized SPTC results for a single vacancy in crystalline graphene. The incremental changes in SPTC values, originating from the defect center and radiating outward are plotted in b). Likewise, c,d) the SPTC and their incremental changes from the defect center in the SW-585 crystalline graphene configuration.

## 4. Results and Discussion

### 4.1. Carbon Structures

#### 4.1.1. Single-Vacancy Defects

The SPTC for hexagonal rings in crystalline graphene are uniformly distributed. However, introducing a single-atom vacancy in an otherwise crystalline graphene structure significantly reduces the TC of nearby atoms. The values for the twofold coordinated carbon atoms near the vacancy are ≈27% that of the threefold coordinated atoms in the bulk region.

As shown in **Figure 3**a, the atoms adjacent to the vacancy (highlighted in blue) exhibit the lowest SPTC. An incremental, radially isotropic increase in normalized SPTC is observed moving outward from the defect center. To quantify anisotropy, the SPTC along two angular directions—0° and 30°, marked by the red and green lines in Figure 3a—is shown in Figure 3b. The variation in SPTC with distance follows a similar trend in both directions.

#### 4.1.2. SW-585 Defects

In the defective SW-585 configuration shown in Figure 3c, the atoms in the five- and eight-membered rings are the lowest thermal conduction active sites, with the normalized SPTC reduced by ≈60% compared to the bulk maximum. The SPTC gradually increases in all directions moving away from the defective region. The changes along two angular directions, 90° and 30°, represented by red and green lines in Figure 3c, are illustrated in Figure 3d. The plot shows an isotropic variation in SPTC, returning to the maximum (bulk) values at around 10 Å away from the defect center.

#### 4.1.3. Amorphous Graphee

In the $aGr_{2000}$ model shown in **Figure 4**(i), atoms with only two neighbors exhibit SPTC values up to 74% of those of three-coordinated atoms. However, the SPTC for the three-coordinated carbon atoms varies depending on their local ring connectivity: atoms in nonhexagonal rings are less thermally active than those in hexagonal rings, similar to what is observed in crystalline graphene. To confirm this, a small 60-atom amorphous graphene model ($aGr_{60}$) was created by extracting a layer from an amorphous graphite model,[19,20] followed by conjugate gradient relaxation while maintaining periodic boundary conditions. Its SPTC analysis, shown in Figure 4(ii), also indicates higher values for hexagonal rings compared to nonhexagonal ones. The ring distribution for both amorphous graphene models is illustrated in Figure 4(iii).

We note a similarity between thermally and electronically conductive regions. Refs. [19,20], which report the electronic transport properties of amorphous graphite, report that in the amorphous graphene layers, the hexagonal carbon rings favor electron flow, while connections between pentagonal and heptagonal rings impede it. A similar observation was made for ring disorder in amorphous carbon nanotubes.[29]

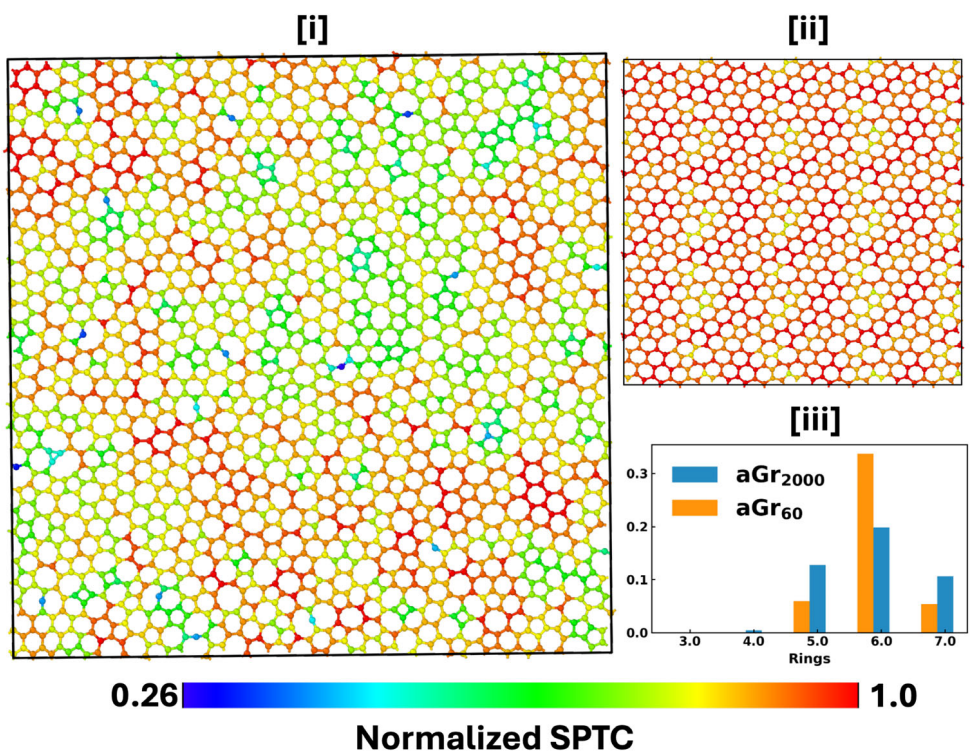

**Figure 4.** Normalized SPTC results for amorphous graphene with i) 2000 atoms ($aGr_{2000}$) and ii) 60 atoms ($aGr_{60}$). iii) The ring distribution for the $aGr_{2000}$ and $aGr_{60}$ models.

## 4.2. Silicon Structures

To test our implementation of Allen and Feldman formulation for the thermal conductivity,[12] we compared the temperature and frequency dependence of TC in amorphous silicon produced by our implementation with the results of Feldman et al.[30] as shown in **Figure 5**a,b. The TC temperature dependence obtained here is consistent with that in ref. [30], while the peak in the frequency dependence curve (the gray dashed line at 0.24 meV) matches the position they reported at 175 K. Note that the consistency is strong despite the use of a different interatomic potential and model in our work.

### 4.2.1. Diamond Silicon and Single-Vacancy Defect

In a defect-free diamond-silicon structure, all atoms contribute equally to the total TC of the system. On the other hand, local TC on atomic sites in the vicinity of the single-vacancy defect is sharply reduced. Such undercoordinated silicon atoms with dangling bonds exhibit ≈70% reduction in SPTC, as compared to the bulk-crystal value. The SPTC evolution with distance from the center of the defect in the (111) plane—along green and red lines shown in **Figure 6**a—is presented in Figure 6b.

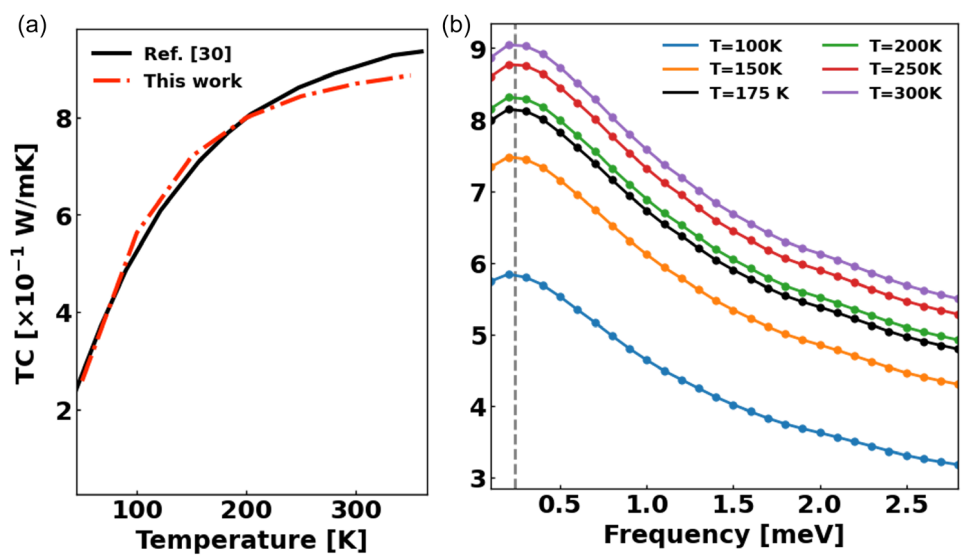

**Figure 5.** a) Temperature dependence of TC for amorphous silicon, with results from this study compared to those of Feldman and co-workers.[30] b) Frequency dependence of TC at various temperatures, with the dashed gray line marking the peak at 175 K. This peak aligns with the corresponding plot at the same temperature in ref. [30].

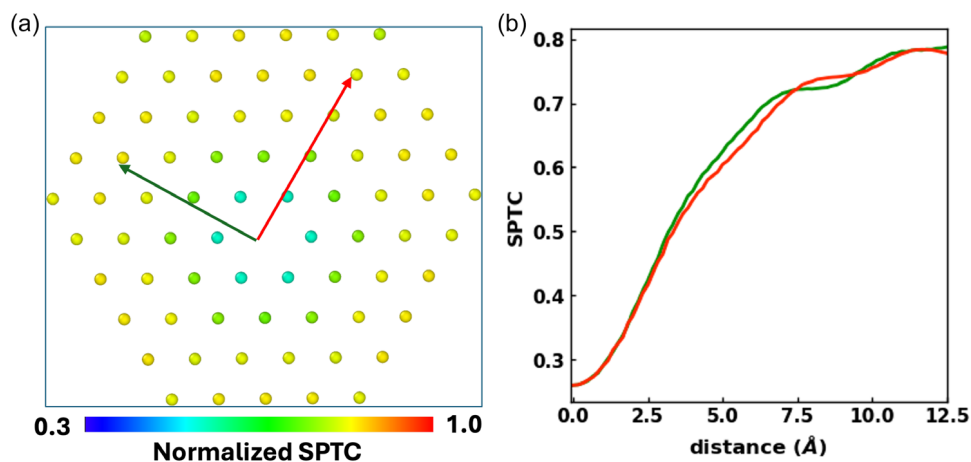

**Figure 6.** a) Normalized SPTC data for a single-vacancy defect in crystalline Si, projected along the (111) plane intersecting the defect site. b) The variation in SPTC with distance from the defect center along the red and green lines shown in panel (a).

### 4.2.2. Twin Grain Boundary in Silicon

In the SPTC analysis of the $\Sigma3(112)/[1\bar{1}0]$ silicon GB in **Figure 7**a, the five-coordinated atoms (floating bonds) at the GB exhibit lower SPTC compared to the bulk silicon and also influence the thermal activity of nearby atomic sites. Furthermore, a strong correlation between bonding environment and SPTC is evident: in Figure 7b, the five-coordinated Si atoms labeled "i" show an SPTC reduction of around 79% compared to the bulk crystalline region labeled "ii". Figure 7c further indicates that bond length distortions at the GB contribute to the reduced SPTC, as atomic pairs with bond lengths exceeding 2.36 Å (the bond length in diamond silicon) exhibit significantly lower SPTC values.

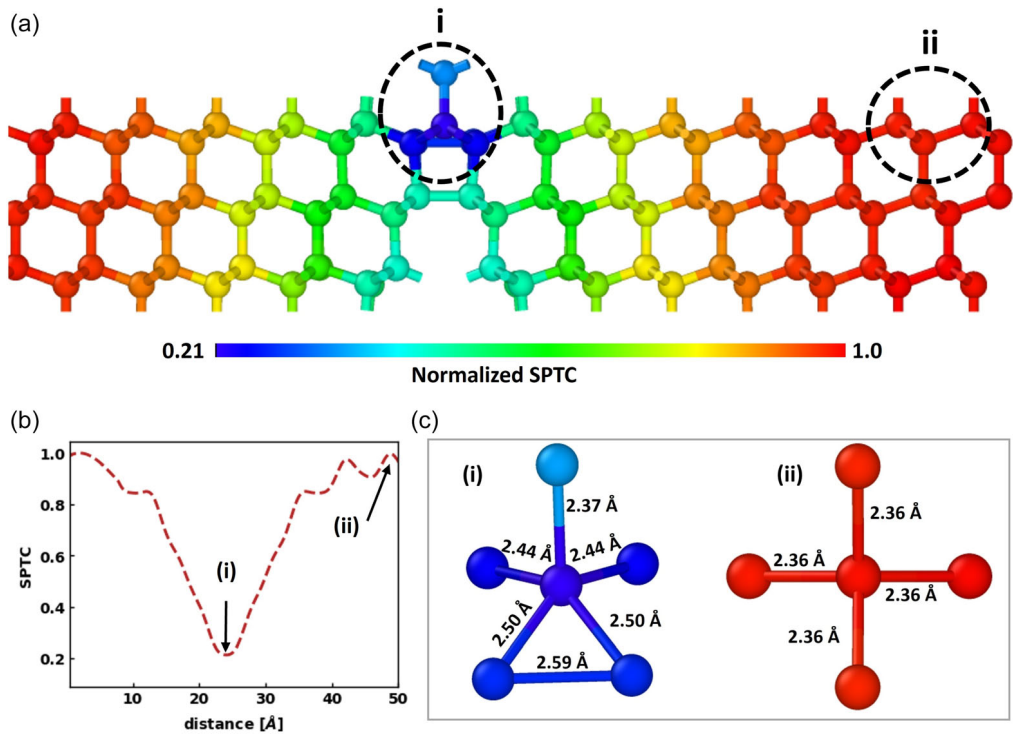

**Figure 7.** a) SPTC for the Si $\Sigma3(112)/[1\bar{1}0]$ twin GB, highlighting the correlation between SPTC values and bond lengths in different Si configurations: i) fivefold coordination at the GB and ii) fourfold coordination at the bulk region. b) The variation in SPTC for atoms along the line labeled "i" and "ii", with their c) bond length distributions.

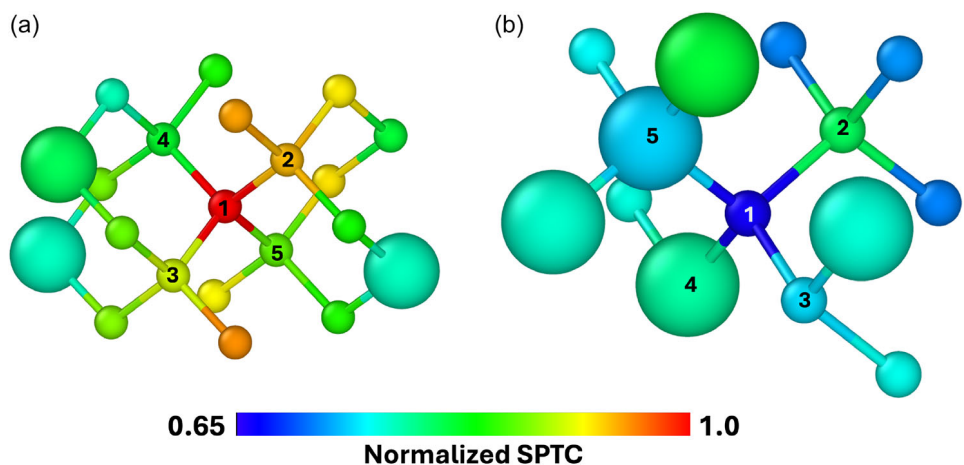

**Figure 8.** Bonding environment for Si atoms with a) the highest and b) the lowest SPTC in the $Si_{0.75}Ge_{0.25}$ crystal. (a) The silicon atom (1) has the highest SPTC and is connected to silicon atoms (2–5), which are also connected to other silicon atoms. (b) The silicon atom (1), which has the lowest SPTC among silicon atoms in the system, is connected to two silicon atoms (2, 3) and two germanium atoms (4, 5).

### 4.3. Mass and Chemical Disorder

The SPTC of the Ge-substituted $Si_{0.75}Ge_{0.25}$ crystal reveals a reduction in TC, consistent with expectations for a system exhibiting mass disorder. The average SPTC for Ge atoms in the $Si_{0.75}Ge_{0.25}$ crystal is ≈7% lower than that for Si atoms. Notably, regions with a higher Si concentration display comparatively higher conductivity than those with a greater Ge concentration. The TC values obtained for the $Si_{1-x}Ge_x$ system are as follows: for $x = 0.25$, the value is $1.50\,W\,mK^{-1}$; for $x = 0.5$, it remains $1.50\,W\,mK^{-1}$; and for $x = 0.75$, it increases to $2.46\,W\,mK^{-1}$. These results align with the trends observed in $Si_{1-x}Ge_x$ superlattices.[31]

Further, atomistic analysis of the $Si_{0.75}Ge_{0.25}$ crystal reveals distinct correlations between SPTC and the local environment of individual atoms. For example, a silicon atom exhibits the highest SPTC when connected to four neighboring silicons, which are also connected with other silicon atoms. Conversely, a silicon atom exhibits the lowest SPTC when bonded to the germanium atoms and/or connected to silicon atoms which are connected to germanium atoms. **Figure 8**, provides a visual description of the thermal characteristics of silicon atoms in different bonding environments within the Si–Ge system.

## 5. Conclusion

This study introduces the SPTC methodology that employs the Green–Kubo theory to map TC onto atomic sites, thus identifying more or less thermally active regions. To illustrate the utility of this approach, we calculated the SPTC to carbon, silicon, and silicon–germanium systems, showing that point, planar (GB), and mass defects in these systems considerably reduce local TC. Although the developed methodology does not account for electronic contributions, limiting quantitative comparisons with experimental data, it effectively predicts thermal conduction paths consistent with expected defect behavior in these materials. To date, we have implemented the method "exactly". However, since the dynamical matrix is short ranged,[32] it may become possible to explore much larger systems.

## Acknowledgements

The authors thank Professor J. J. Dong for helpful discussions. This research was supported by ONR grant N00014-23-1-2773. It also used computational resources from the Advanced Cyberinfrastructure Coordination Ecosystem: Services & Support (ACCESS) program, through allocation DMR190002, MAT240030, funded by the U.S. National Science Foundation (NSF) grant nos. 2138259, 2138286, 2138307, 2137603, and 2138296.

## Conflict of Interest

The authors declare no conflict of interest.

## Author Contributions

**Aashish Gautam**: Data curation (equal); Formal analysis (equal); Investigation (equal); Methodology (equal); Software (equal); Visualization (equal); Writing—original draft (equal); Writing—review & editing (equal). **Yoon Gyu Lee**: Conceptualization (supporting); Data curation (equal); Formal analysis (supporting); Investigation (supporting); Methodology (equal); Software (lead); Validation (equal); Visualization (supporting); Writing—original draft (equal); Writing—review & editing (supporting). **Chinonso Ugwumadu**: Conceptualization (supporting); Data curation (supporting); Formal analysis (equal); Investigation (equal); Methodology (equal); Software (equal); Supervision (supporting); Validation (equal); Visualization (equal); Writing—original draft (supporting); Writing—review & editing (equal). **Kishor Nepal**: Methodology (supporting); Software (supporting); Visualization (equal); Writing—original draft (equal); Writing—review & editing (equal). **Serge Nakhmanson**: Formal analysis (equal); Funding acquisition (equal); Project administration (equal); Resources (equal); Validation (equal); Writing—review & editing (equal). **David A. Drabold**: Conceptualization (lead); Data curation (lead); Formal analysis (lead); Funding acquisition (equal); Investigation (lead); Methodology (lead); Project administration (equal); Resources (lead); Software (supporting); Supervision (lead); Validation (lead); Visualization (lead); Writing—original draft (lead); Writing—review & editing (lead).

## Data Availability Statement

The data that support the findings of this study are available from the corresponding author upon reasonable request.

[17] J. Tersoff, *Phys. Rev. B* **1989**, *39*, 5566.
[18] A. Stone, D. Wales, *Chem. Phys. Lett.* **1986**, *128*, 501.
[19] R. Thapa, C. Ugwumadu, K. Nepal, J. Trembly, D. A. Drabold, *Phys. Rev. Lett.* **2022**, *128*, 236402.
[20] C. Ugwumadu, R. Thapa, K. Nepal, D. A. Drabold, *Eur. J. Glass Sci. Technol., Part B* **2023**, *64*, 16.
[21] S. Nosé, *Mol. Phys.* **1984**, *52*, 255.
[22] W. G. Hoover, *Phys. Rev. A* **1985**, *31*, 1695.
[23] W. Shinoda, M. Shiga, M. Mikami, *Phys. Rev. B* **2004**, *69*, 134103.
[24] V. Kapko, D. A. Drabold, M. F. Thorpe, *Phys. Status Solidi B* **2010**, *247*, 1197.
[25] Y. Li, F. Inam, A. Kumar, M. F. Thorpe, D. A. Drabold, *Phys. Status Solidi B* **2011**, *248*, 2082.
[26] A. Jain, S. P. Ong, G. Hautier, W. Chen, W. D. Richards, S. Dacek, S. Cholia, D. Gunter, D. Skinner, G. Ceder, K. A. Persson, *APL Mater.* **2013**, *1*, 011002.
[27] B. R. Djordjević, M. F. Thorpe, F. Wooten, *Phys. Rev. B* **1995**, *52*, 5685.
[28] S. Fujii, A. Seko, *Comput. Mater. Sci.* **2022**, *204*, 111137.
[29] C. Ugwumadu, R. Thapa, Y. Al-Majali, J. Trembly, D. A. Drabold, *Phys. Status Solidi B* **2023**, *260*, 2200527.
[30] J. L. Feldman, M. D. Kluge, P. B. Allen, F. Wooten, *Phys. Rev. B* **1993**, *48*, 12589.
[31] Y. Lee, A. J. Pak, G. S. Hwang, *Phys. Chem. Chem. Phys.* **2016**, *18*, 19544.
[32] P. Ordejón, D. A. Drabold, R. M. Martin, S. Itoh, *Phys. Rev. Lett.* **1995**, *75*, 1324.
*Phys. Status Solidi RRL* **2024**, 2400306

2400306 (9 of 9)

© 2024 The Author(s). physica status solidi (RRL) Rapid Research Letters published by Wiley-VCH GmbH